# On the permeability of fractal tube bundles

I. Zinovik, and D. Poulikakos

Laboratory of Thermodynamics in Emerging Technologies, Department of Mechanical and Process Engineering, ETH Zurich, Zurich 8092, Switzerland

izinovik@yahoo.com  Ph: +41 44 632 2435    Fax: +41 44 632 1176

**Abstract**

The permeability of a porous medium is strongly affected by its local geometry and connectivity, the size distribution of the solid inclusions and the pores available for flow. Since direct measurements of the permeability are time consuming and require experiments that are not always possible, the reliable theoretical assessment of the permeability based on the medium structural characteristics alone is of importance. When the porosity approaches unity, the permeability-porosity relationships represented by the Kozeny-Carman equations and Archie's law predict that permeability tends to infinity and thus they yield unrealistic results if specific area of the porous media does not tend to zero.

The goal of this paper is an evaluation of the relationships between porosity and permeability for a set of fractal models with porosity approaching unity and a finite permeability. It is shown that the tube bundles generated by finite iterations of the corresponding geometric fractals can be used to model porous media where the permeability-porosity relationships are derived analytically. Several examples of the tube bundles are constructed and relevance of the derived permeability-porosity relationships is discussed in connection with the permeability measurements of highly porous metal foams reported in the literature.

**Introduction**

Fractals are proved to be a useful model of porous media with complex structure. The results of application of fractal models to analysis of flow in porous structures are covered in extensive reviews (Gimenz et al. 1997, Perfect et al. 2006, Perrier et al. 2000, Yu and Liu 2004). Such studies are often focused on the link between permeability and the fractal dimension of porous structures. A systematic study of this topic begins with the analysis given by Jacquin and Adler (1987), where a porous medium is modeled as a set of parallel solid rods with an axial Stokes flow around the solid. The cross section of the medium is represented by a prefractal of the Sierpinski carpet shown in Fig. 1 (white squares depict the solid inclusions). An extensive series of numerical simulations of the flow is conducted for different prefractals with the porosity decreasing to zero. The results show that the permeability can be approximated by a power-law function of the porosity if the power coefficient equals (4-D)/(2-D) where D is the dimension of the fractal set. This power law approximation of the permeability of a fractal medium as a function of porosity can be seen as an extension of the Kozeny-Carman equation and is called Archie's law (Gimenez et al. 1997). Further investigations indicate that for low-porosity media, Archie's law exponent can either be defined as a function of the fractal dimension or serves as a fitting parameter (Henderson et al. 2010) of empirical correlations. In these equations, the permeability is inversely proportional to one minus the porosity and thus tends to infinity when the porosity approaches unity (Henderson et al. 2010, Costa 2006).

Most of the studies of fractal models of permeability are focused on flow in consolidated soils and fractured rock where the porosity of the media approaches zero. However, there exist many applications in engineering where the limit at the other end of the porosity spectrum (porosity approaching unity) is at play. To this end, in many engineering applications involving metal foams (Boomsma, and Poulikakos 2001, Boomsma et al. 2003, Boomsma et al. 2003a), aerogels (Pierre and Pajonk 2002, Donatti et al. 2005, Gurav et al. 2010), bone matrices (Predoi-Racila et al. 2010, Fritton et al. 2009), scaffolds for tissue engineering (Deville et al. 2006, O'Brien et al. 2007, Hirata 2010), bundles of carbon nanotubes (Neimark et al. 2003) filters and catalytic reactors (Thormann et al. 2007) permeability relations for high porosity media are necessary. Since many of the high porosity materials also possess a complex hierarchical structure, the objective of this work is to propose fractal models for the finite permeability of porous media when the porosity tends to unity.

The model porous medium studied herein is a bundle of parallel tubes with the cross section of the bundle being specified as a two-dimensional prefractal. For example, the inversion of pore and solid spaces of the Sierpinski carpet shown in Fig. 1 can be seen as the cross section of a bundle of parallel tubes with square cross sections. If the number of iterations for the prefractal construction tends to infinity, the porosity of the bundles tends to unity and the model porous medium is as a bundle of an infinite number of tubes with infinitely thin walls. In hydrology, the fractal models are applied to analysis of permeability of soils (Perfect et al. 2006). In the mathematical studies, similar fractals are sometimes called the sprays and their structural properties are expressed in terms of the Riemann zeta-function (Lapidus 2010).

It is clear that the bundle of tubes with infinitely thin walls have a finite permeability since the tube walls impose a resistance on the flow. Thus, the permeability equations (Henderson et al. 2010, Costa 2006) derived for low-porosity media will fail to describe permeability of the prefractal tube bundles, since the permeability of these bundles should converge to a finite value defined by the corresponding fractal. The goal of this study is the derivation and evaluation of relationships between porosity and permeability for a set of fractal models with porosity approaching unity and a finite permeability.

**Section 1. Permeability of prefractal tube bundle with power law distribution of tube diameters**

First we will derive the relation between porosity and permeability of a prefractal tube bundle making assumptions usual for the analysis of the Darcy permeability of two-dimensional fractal porous media. We assume that the prefractal bundle consists of parallel tubes of a fixed length $L$ with circular cross sections, and the flow in all tubes obeys the Hagen-Poiseuille equation and is driven by the same pressure gradient $\Delta P/L$. In addition to this, we assume that the distribution of the hydraulic diameters $d$ of the bundle tubes is described by a power law.

For physical porous media, the last assumption about power law distribution does not hold exactly but serves as an approximation of actual data. A geometric example of fractal sets that satisfy this assumption asymptotically is the Apollonian gasket (see Fig. 2). Numerical calculations (Boyd 1982, Manna and Herrmann 1991) indicate that, after a certain number of iterations, the distribution function converges to a power law with the exponent that becomes the dimension of the Apollonian fractal. The pre-factor of the power law in the case of integral Apollonian packing is numerically calculated by Fuchs and Sanden (2009).

The assumption of the Hagen-Poiseuille regime of the flow implies that the flow rate $q$ in a tube with a hydraulic diameter $d$ can be calculated as follows: $q(d)=\alpha d^4$, where $\alpha=(\beta/\mu)(\Delta P/L)$, $\mu$ is the dynamic viscosity of the fluid, and $\beta$ is a constant form factor for circular tubes. The number density of the distribution of the hydraulic diameters $d$ of the bundle tubes can be written as follows: $n(d)=C d^{-\tau}$, where C is a constant with the dimension of length in power of $\tau-1$ and the exponent $\tau$ is related to the fractal dimension $D$ via $\tau=1+D$. Using the introduced notation, the tube cumulative number distribution (the number of tubes with diameter smaller than $d$) $N$, the porosity $\varphi$, and the total flow rate $Q$ in the tube bundle can be written as normalized corresponding moments $m_k$ of the distribution density function $n(d)$:

$$N = m_0 = \int_d^{d_0} n(\xi)d\xi = \frac{C}{D}(d^{-D} - d_0^{-D}) \qquad (1)$$

$$\varphi = \frac{\pi}{4A} m_2 = \frac{\pi}{4A}\int_d^{d_0} n(\xi)\xi^2 d\xi = \frac{\pi C}{4A(2-D)}(d_0^{2-D} - d^{2-D}) \qquad (2)$$

$$Q = \alpha m_4 = \int_d^{d_0} n(\xi)q(\xi)d\xi = \alpha \int_d^{d_0} n(\xi)\xi^4 d\xi = \frac{\alpha C}{4-D}(d_0^{4-D} - d^{4-D}) \qquad (3)$$

In the Eqs (1-3), $d_0$ and $d$ are the maximum and minimum hydraulic diameters in the prefractal bundle, respectively, after a certain number of iterations which generate the bundle cross section. The cross sectional area of the bundle is denoted by $A$. If the number of iterations tends to infinity, the minimum diameter tends to zero while the porosity tends to unity. Thus, Eq. 2 is reduced to the following equation:

$$1 = \frac{\pi C}{4A(2-D)} d_0^{2-D}, \qquad (4)$$

and the total flow rate in the fractal tube bundle $Q_f$ is obtained from the Eq. 3 when $d=0$ as follows

$$Q_f = \frac{\alpha C}{4-D} d_0^{4-D} \tag{5}$$

The minimum diameter $d$ can be expressed as a function of the porosity using Eq. 2. Substitution of the resulting equation into Eq. 3 and elimination of the minimum and maximum diameters from the resulting expression with the help of Eq. 4-5 leads to the following equation for the total flow rate:

$$Q = Q_f(1 - (1-\varphi)^{\frac{4-D}{2-D}}). \tag{6}$$

The Darcy permeability $K_D$ of the bundle can be written using the Darcy law:

$$K_D = \frac{\mu}{\Delta P/L} \frac{Q}{A}, \tag{7}$$

Substitution of Eq. 6 into Eq. 7 yields the final equation connecting the Darcy permeability and porosity of the prefractal porous medium:

$$K_D = K_{Df}(1 - (1-\varphi)^{\frac{4-D}{2-D}}), \tag{8}$$

where the permeability of the fractal tube bundle $K_{Df}$ is:

$$K_{Df} = \frac{C\beta}{(4-D)A} d_0^{4-D} \tag{9}$$

The derivation of Eq. 8 can be repeated for an arbitrary shape of tube cross section if the hydraulic diameter $d$ is replaced by a characteristic length scale which equals to square root of the tube cross section area and form factor $\beta$ is replaced by a form factor specific to the cross section of the tubes. A collection of the corresponding form factors for various tube cross sections can be found in (Shah 1978). Note that the permeability equation of the prefractal tube bundle has the same exponent coefficient as the coefficient which is obtained in (Jacquin and Adler 1987) based on numerical simulations of the flow in a low porosity fractal porous medium. In case of the fractal tube bundle, the specific form of the exponent coefficient in the permeability-porosity equation 8 is due to the algebraic link between the second and the forth-order moments (Eq. 2 and 3) of the power-law distribution function $n(d)$.

**Section 2. Geometric fractals**

Since the assumption of the power law distribution of pore sizes is only an approximation, a convenient way to evaluate the permeability-porosity in actual prefractals is to utilize the geometric fractals as generating sets of the tube bundle cross section. To exemplify the derivation of the permeability-porosity relation in geometric prefractal tube bundles, we consider the Sierpinski carpet shown of Fig. 1. The following notation will be used to characterize the fractal structure: 1) the area of the bundle is denoted by $A$; 2) the number of tubes that are added into the bundle at iteration "$k$" is denoted as $N_k$; 3) the area of cross section of every tube which is added at the iteration "$k$" is $a_k$; 4) a pattern (tube number) scaling factor $x$ is defined as $x=N_{k+1}/N_k$; and 5) a length scaling factor $y$ is defined as $y=(a_k/a_{k+1})^{1/2}$. For the square Sierpinski carpet, $N_1=1$, $a_1=A/9$ and the factors $x$ and $y$ equal 8 and 3 respectively. With this notation, the porosity of the fractal tube bundle can be written as a geometric series:

$$\varphi_f = \frac{1}{A}(N_1 a_1 + N_2 a_2 + N_3 a_3 + \cdots) = \frac{1}{y^2} + \frac{x}{y^4}\sum_{k=0}^{\infty}\left(\frac{x}{y^2}\right)^k = \frac{1}{y^2-x} \qquad (10)$$

Note that since $\varphi_f=1$, the scaling factors have to satisfy the equation:

$$y^2-x=1 \qquad (11)$$

The equation is true if $x<y^2$, which is implicitly satisfied in the applied fractals and it can be written as $y^2 - x = N_1$.

As in the previous section, we assume that the flow rates in the tubes can be calculated as $q_k = \alpha\, a_k^2$, with an adequate value of the coefficient $\alpha$. The total flow rate in the fractal tube bundle can be written as a geometric series similar to one shown in Eq. 10 but with squared denominators:

$$Q_f = \sum_{k=1}^{\infty} q_k = \sum_{k=1}^{\infty} \alpha N_k a_k^2 = \alpha A^2 \left(\frac{1}{y^4} + \frac{x}{y^8}\sum_{k=0}^{\infty}\left(\frac{x}{y^4}\right)^k\right) = \alpha A^2 \frac{1}{y^4 - x} \quad (12)$$

Substitution of Eq. 12 into Darcy's law (Eq. 7), yields the Darcy permeability coefficient for the fractal tube bundle $K_{Df}$:

$$K_{Df} = \beta A \frac{1}{y^4 - x} \quad (13)$$

To obtain the relation between permeability and porosity for the prefractal tube bundle, we apply the formula of the sum of the first $n$ terms of a geometric series to Eq. 10 and 12:

$$\varphi = \frac{1}{A}(N_1 a_1 + \cdots + N_n a_n) = \frac{1}{y^2} + \frac{x}{y^4}\sum_{k=0}^{n}\left(\frac{x}{y^2}\right)^k = \frac{1}{y^2} + \frac{x}{y^4}\frac{1-\left(\frac{x}{y^2}\right)^{n+1}}{1-\frac{x}{y^2}} \quad (14)$$

$$Q = \sum_{k=1}^{n} q_k = \alpha A^2\left(\frac{1}{y^4} + \frac{x}{y^8}\sum_{k=0}^{n}\left(\frac{x}{y^4}\right)^k\right) = \alpha A^2\left[\frac{1}{y^4} + \frac{x}{y^8}\frac{1-\left(\frac{x}{y^4}\right)^{n+1}}{1-\frac{x}{y^4}}\right] \quad (15)$$

Note that the pore size distribution function can be calculated using Eq.14 with the coefficient $y$ set to unity and expressed as a parametric function of the power of inverse size $(1/y)^n$ with the iteration number $n$ serving as the parameter of the function.

$$.N = N_1 + \cdots + N_n = 1 + x\sum_{k=0}^{n} x^k = 1 + x\frac{1-x^{n+1}}{1-x} \quad (16)$$

To simplify the notation, we introduce the dimensionless permeabilities $K_f$ and $K$ of the fractal, and prefractal bundles, respectively:

$$K_f = \frac{K_{Df}}{\beta A}; \quad K = \frac{K_D}{\beta A} \tag{17}$$

The denominator in the defined dimensionless permeability refers to the permeability of a single tube with the same cross section as the tube bundle and thus the value of the dimensionless is always less than unity. In the geometric fractals, the scaling factors $x$ and $y$ are related to the fractal dimension as

$$x = y^D \tag{18}$$

Equations 14 and 15 can be re-written substituting Eq. 18 for $x$ and then simplified using Eq.11. The results read:

$$\varphi = y^{-2} + y^{(D-2)}(1 - y^{(D-2)(n+1)}) \tag{19}$$

$$K = y^{-4} + K_f y^{(D-4)}(1 - y^{(D-4)(n+1)}) \tag{20}$$

Eliminating the scaling factor $y^{(n+1)}$ from Eq. 19 and 20 yields the relation between permeability and porosity:

$$K = K_f - K_*(1-\varphi)^{(4-D)/(2-D)}, \tag{21}$$

where

$$K_* = \left[K_f - \frac{1}{y^4}\right]\left[\frac{y^2}{y^2-1}\right]^{(4-D)/(2-D)}. \tag{22}$$

Equations 8 and 21 have identical exponent coefficients thus, when the porosity tends to unity, the permeability of the geometric prefractal behaves the same way as the permeability of a prefractal with exact power law distribution. The distribution function $N=f(1/y^n)$ of the prefractal defined by Eq. 16 can also be approximated by a power law with a high accuracy: a power law curve with an exponent which equals the fractal dimension of the Sierpinski carpet approximates the fractal distribution function with a determination coefficient $R^2 > 0.999$. On the other hand, the functional relation between permeability and porosity of the geometrical prefractal requires not only the fractal permeability $K_f$ as in the case of Eq. 8 but also an additional parameter $K_*$ which depends on the fractal dimension.

Note that Eq. 21 does not depend on the form factor β. The permeability of prefractal bundles of the Sierpinski triangular carpet (Fig. A1) is also defined by Eq. 21 with the corresponding substitutions of the scaling factors: $x=3$, and $y=2$. Three more examples of the permeability-porosity relations for the geometric prefractal tube bundles are presented in the Appendix A. It is shown that such relations can be written in the form of Eq. 21 for the Pinwheel (Radin 1994), the Pascal triangle modulo 5 (Wolfram 1984), and the Koch snowflake prefractals. The permeability-porosity curves of the studied geometric tube bundles together with the curve for the Apollonian gasket are plotted in the Fig. 3. The data points for the Apollonian gasket are generated using a MATLAB script available on the website of company MathWorks which develops mathematical software (Jacquenot 2007).

The tube cross section area decreases rapidly as a function of the iteration number, and the permeability of fractals is approximately equal to the permeability of prefractals after a couple of iterations (see Fig. 3). The shape of the permeability-porosity curve of the Apollonian prefractal (closed triangles in Fig. 3) resembles the curves of the other geometric prefractals. In Fig. 3, the dashed curve fits permeability of the Apollonian prefractal using Eq. 21 where coefficients $K_f$, $K_*$, and *D* are treated as fitting parameters. The fitted value of the coefficient *D* was found to be 0.70, that is, almost two times smaller than the Apollonian fractal dimension. This deviation is due to the fact that at the initial iterations, the size distribution function of the Apollonian gasket deviates significantly from the power law that asymptotically approximates the fractal dimension of the gasket (Manna and H J Herrmann, 1991).

The plots of the pore size distribution functions (see Eq. 16) of two geometric prefractal bundles and the Apollonian gasket are shown in Fig. 4. The solid lines depict fitting which uses power law function with a constant exponent. In case of the geometric prefractal bundles, fitting exponent recovers fractal dimensions of the corresponding generating fractals with accuracy better than 1%. For the Apollonian fractal, fitting exponent was more than 30% smaller than the fractal dimension.

Figure 3 and 4 show that Eq. 21 can serve as a fitting formula for approximation of permeability of both prefractal geometric bundles and the tube bundle generated by the Apollonian gasket. The fitting value of the exponent coefficient *D* recovered the corresponding fractal dimensions in the case of the geometric prefractal bundles but was much smaller in the case of the Apollonian fractal.

In the next section, we construct examples of composite tube bundles and show that the permeability-porosity relationships of these bundles differ from the equation for the geometric prefractals derived above. We also demonstrate that Eq. 21 still can be used as a fitting equation which approximates permeability of the composite bundles.

**Section 3. Permeability of composite (multiple) geometric tube bundles**

We define the double Sierpinski tube bundle as an ordinary Sierpinski bundle discussed above where every tube of the original bundle is filled with the corresponding Sierpinski bundle of the smaller size. Since the Hausdorff dimension of countable unions of sets with the same dimension equals the set dimension, countable unions of the geometric fractals have the same dimension as the original fractal. The double Sierpinski bundle is then a countable set of the Sierpinski bundles and the dimension of the fractal cross section of the double carpet is the same as the dimension of the original Sierpinski carpet. Thus the pore size distribution function of the double fractal can be approximated with a high accuracy by a power law with the same exponent coefficient as the dimension of the original fractal. On the other hand, the permeability of the double fractal is lower than that of the original fractal. For example, the dimensionless permeability of double Sierpinski bundle, equals the square of the permeability of the original bundle (see Appendix B).

The iterative procedure mentioned above can also be used to construct two different composite fractal bundles using a pair of different geometric fractals. For example, if the composite bundle is generated from the Sierpinski triangular fractal filled with the Pascal triangular modulo 5 fractal, the fractal dimension of the composite bundle equals the dimension of the Pascal fractal and vice versa: if the bundle consists of the Sierpinski triangular carpets filling the Pascal triangular fractal, the bundle dimension is the same as the fractal dimension of the Sierpinski carpet. Despite the different fractal dimensions, both composite fractal bundles have the same dimensionless permeability (see Appendix B). In both cases, the permeability of the bundle equals the permeability of the Sierpinski carpet times the permeability of the Pascal triangle modulo 5 fractal.

For the prefractal composite bundles, permeability, porosity, and pore size distribution function can be expressed in terms of the partial sums of corresponding geometric series in the same manner as for the geometric prefractals discussed in the previous sections. For example, the equation for permeability of the double Pinwheel prefractal can be written as follows (see Appendix C):

$$K = k_1 + k_2(k_3 + k_4(k_5 + k_6\sqrt{k_7 + k_8\varphi})^{\frac{4-D}{2-D}})^2 \qquad (23)$$

where $k_i$, $i=1$-$8$ are rational functions of the scaling factors $x$ and $y$.

The example shows that the permeability-porosity formula of the composite fractal bundles differs from Eq. 21 derived for the geometric fractal bundles.

Despite the difference between Eq. 21 and Eq. 23, the plot of permeability of the double Pinwheel prefractal resembles the permeability curves of the geometric prefractal bundles shown in Fig. 3. It was found that Eq. 21 can accurately ($R^2>0.99$) approximate the permeability of the double Pinwheel tube bundle, and double Sierpinski tube bundles defined above. For the approximation, all three coefficients of Eq. 21 have to be treated as adjustable positive parameters.

In all three cases, the fitted value of coefficient $D$ was greater than the corresponding fractal dimension of the bundle generated gaskets: 1.88 vs. 1.72 for the double Pinwheel fractal, 1.79 vs. 1.59 for the double Sierpinski triangular carpet, and 1.89 vs. 1.97 for the double Sierpinski rectangular carpet. The examples show that fitting values of coefficient $D$ cannot be used to estimate fractal dimension of the bundle generating fractals. Nevertheless Eq. 21 can serve as a useful fitting formula approximating permeability of the high porosity tube bundles.

**Discussion**

The application of the capillary model of flow to actual three-dimensional porous media has apparent limitations. Nevertheless, this model is often used to provide a basis for discussion of possible factors influencing the flow in porous media. Since the porosity of the prefractal tube bundles tends to unity, it is useful to test if this model could serve as a tool for approximation of permeability-porosity relationships of highly porous media. One example of a high porosity medium is the metal foams which are often studied in connection with heat and mass transfer processes (Boomsma, and Poulikakos 2001, Boomsma et al. 2003, Boomsma et al. 2003a).

Difficulties of estimation of the permeability of metal foams sometime lead to counterintuitive assumptions which are not supported by the experimental results. For example, semi-empirical fitting suggested in (Bhattacharya et al. 2002) for the measured permeability is based on a definition of tortuosity which predicts that the high porosity limit of the tortuosity is not unity (see Eq. 20 in Bhattacharya et al. 2002). Such definition of tortuosity contradicts most of the experimental results reported in the literature where tortuosity is approximated as a power of the porosity.

In Fig. 5, the permeability values from Table 2 of Bhattacharya et al. 2002 are plotted as a function of porosity. The data shown in Fig. 5 does not favor the conclusion of an infinite permeability increase which would be predicted by the Archie's law in high porosity limit over the possibility of gradual leveling of the permeability as it follows from the prefractal bundle model. Moreover, the permeability of the foam with 40 pores per inch (closed symbols 4 in Fig. 5) remains approximately the same for all tested porosities as it would be according to the permeability-porosity formula for the prefractal bundles when the porosity approaches unity.

A possible explanation of the qualitatively similar behavior of the permeability of the metal foams shown in Fig. 5 and the prefractal tube bundles can be found in the following example: let us assume that a sample of a solid mesh is inserted into a tube with flowing fluid. If the sample is relatively small and the mesh pores are relatively large, some space will be available for flow through the sample along almost straight flow lines. This space can be visualized by a projection of the mesh on the plane perpendicular to the tube axis. This projection may look like a cross section of a tube bundle composed of the tubes with different hydraulic diameters (see illustrative example in Appendix F). If the space available for flow along straight or slightly curved flow lines has size distribution function similar to the prefractal bundles, the permeability-porosity relationships of such foams should be similar to the permeability formulas derived herein for the prefractal bundles.

The permeability-porosity relation defined by Eq. 21 was used as a fitting equation for approximation of permeability data for 5 ppi, 10 ppi and 20 ppi foam which is shown in Fig. 5. This data was also fitted by the least square method utilizing the Archie's law formula $K = p\, \varphi^r/(1-\varphi)$ for low porosity fractals proposed in (Costa 2006) and an empirical correlation $K = p\, \varphi^{r+1}/(1-\varphi)^r$ (Rodriguez et al. 2004) developed for chaotically oriented fibers where m and n are fitting parameters. In all cases but one (dashed line in Fig. 5), the prefractal tube bundle formula provided a better fit i.e. smaller root mean square deviation from the data. The regression coefficients are shown in Appendix E.

Since the number of data points in Fig. 5 is limited, applicability of Eq. 21 as an accurate fitting formula needs further experimental investigations. Nevertheless, it is clear that fitting permeability with the Archie's law will be less accurate than with Eq. 21 if the permeability in high porosity limit becomes almost constant as it is for the foam with 40 ppi in Fig. 5.

Derivation of the permeability equations of prefractal tube bundles can also be used to provide a guess when permeability curves of Eq. 8 or Eq. 21 should serve as a better approximation than the Archie's law. For the derivation, it was assumed that the prefractal bundles are constructed by iterative addition of the tubes with gradually decreasing diameters. All steps of the derivation of these permeability equations can also be repeated assuming that the iterations add tubes with gradually increasing diameters. For example the lower limit of the integral in Eq.2 in this case will be zero and the upper limit will denote a maximum tube diameter which increases at every iteration. The permeability equations Eq. 8 and Eq. 21 will then take form $K_D = K_{Df}\, \varphi^{\frac{4-D}{2-D}}$ which is asymptotically equivalent to the Archie's law when the porosity tends to zero. Thus one may guess that the Archie's law provides a better approximation if the permeability of a porous medium increases due to addition of the pores with gradually increasing diameters while fitting curves of Eq. 21 could be used if increase of porosity of the medium is caused by rise of the pore fractions with consequently decreasing diameters.

The different concavity of curves of the Archie's law and Eq. 21 could also be discussed in the context of permeability change due to clogging of the pores or internal erosion in the porous matrix. For example, if increase of the flow rate cases erosion of fines from the pore fractions with greater diameters, change of the porosity is due to increasing number of bigger pores and thus the permeability may follow a concave curve similar to the curves described by the Archie's law (see permeability data measured during erosion in Fig. 8 of Scheuermann 2010). On the other hand, decrease of the flow volume may not only stop the erosion but also lead to deposition of particles suspended in the flow and thus to a decrease of the porosity. If this decrease of porosity is caused by clogging of the pore fractions with smaller diameters, the permeability change may follow a curve described by Eq. 21 with concavity opposite to the concavity of curves of the Archie's law (see permeability data measured during particle deposition in Fig. 6 of Mays 2005). If these two trends in permeability would be observed in the same porous medium during cyclic changes of the flow rate, the curves of Eq. 21 and the Archie's law could serve as approximations of upper and lower parts of the curve of hysteresis of the hydraulic permeability with flow rate.

**Conclusion**

Prefractal tube bundles generated by finite iterations of the corresponding geometric fractals can be used as a model porous medium where permeability-porosity relationships are derived analytically as explicit algebraic equations. Unlike the equations of the Kozeny-Carman and Archie's laws, the permeability-porosity equations of the prefractals predict a finite permeability when the porosity approaches unity. It is shown that the model of prefractal tube bundles can be used to obtain fitting curves of the permeability of high porosity metal foams and to provide insight on permeability-porosity correlations of the capillary model of porous media.

**Acknowledgements**

The authors are indebted to anonymous reviewers of Journal of Transport in Porous Media for corrections and suggestions improved this manuscript.

**Appendix**

A) Permeability of geometric fractal bundles

The Pinwheel fractal (Fig. A2) has the scaling factors $x=4$ and $y=5^{1/2}$ and the fractal dimension $D=log(x)/log(y)=1.7227$. The porosity and permeability of the fractal can be written as follows:

$$\varphi_f = \frac{1}{y^2}\sum_{k=0}^{\infty}\left(\frac{x}{y^2}\right)^k = \frac{1}{y^2-x} = 1 \quad \text{(A1)}$$

$$K_f = \frac{1}{y^4}\sum_{k=0}^{\infty}\left(\frac{x}{y^4}\right)^k = \frac{1}{y^4-x} \quad \text{(A2)}$$

The permeability-porosity is obtained to be:

$$K = K_f\left(1-(1-\varphi)^{\frac{4-D}{2-D}}\right) \quad \text{(A3)}$$

Note that for the Pinwheel prefractal, the coefficient $K_*=1$ thus the permeability-porosity relation of this prefractal has the same form as for a prefractal with the power law distribution defined by Eq. 8.

The Pascal triangle modulo 5 fractal (Fig. A3) has the scaling factors $x=15$, $y=5$ and the fractal dimension $D=1.6826$ respectively. The porosity and permeability of the fractal can be written as follows:

$$\varphi_f = \frac{z}{y^2} + \frac{z}{x}\sum_{k=2}^{\infty}\left(\frac{x}{y^2}\right)^k = z\frac{1}{y^2-x} = 1 \quad \text{(A4)}$$

$$K_f = \frac{z}{y^4} + \frac{z}{x}\sum_{k=2}^{\infty}\left(\frac{x}{y^4}\right)^k = z\frac{1}{y^4-x}, \quad \text{(A5)}$$

where an additional common multiple $z=10$ which is the number $N_1$ of triangles which are introduced at the first iteration of the construction of the fractal. Thus one may note that for the geometric fractals considered above, $y^2-x=N_1$. The permeability-porosity relation is defined by the following equation:

$$K = K_f(1 - x(\frac{y^2}{z}(1-\varphi))^{\frac{4-D}{2-D}})  \qquad (A6)$$

The Koch snowflake fractal has the scaling factors $x=4$, $y=3$ and the fractal dimension $D=1.2619$ respectively. We define a Koch tube bundle as the bundle composed of tubes with triangular (equilateral) cross sections and for convenience of notation, we subdivide initial triangle into 9 smaller triangles. Thus we assume that before the first iteration, the bundle already has 9 tubes which cross sections are defined by these 9 triangles. The porosity and permeability of the fractal Koch bundle can be written as follows:

$$\varphi_f = \frac{5}{8}(1 + \frac{z}{x}\sum_{k=1}^{\infty}(\frac{x}{y^2})^k) = \frac{5}{8}(1 + z\frac{1}{y^2-x}) = 1 \qquad (A7)$$

$$K_f = \frac{25}{64}(\frac{1}{9} + \frac{z}{x}\sum_{k=1}^{\infty}(\frac{x}{y^4})^k) = \frac{25}{64}(\frac{1}{9} + z\frac{1}{y^4-x}), \qquad (A8)$$

where and additional factor $z=3$. Factor z equals to $N_1$, the number of triangles which are introduced at the first iteration of the construction of the Koch fractal, i.e. the three smaller triangles which are attached to the sides of the initial triangle at the first iteration (see Fig. A4). For this fractal, the last equality of Eq. A7 can be re-written as $y^2-x=(5/3)N_1$. Additional multiplier 5/3 appears as a result of assumptions about initial steps of construction of the bundle. While for the other cases, porosity before first iteration equals zero, construction of the Koch bundle is defined such a way that the bundle is composed of 9 tubes even before the iterations.

The permeability-porosity relation of the Koch tube bundle is:

$$K = K_f - \frac{zy^4}{x(y^4-x)} \left(\frac{8}{5}\frac{x(y^2-x)}{y^2 z}(1-\varphi)\right)^{\frac{4-D}{2-D}} \qquad (A9)$$

B) Permeability of composite fractal bundles.

We denote flow rate of the Sierpinski bundle with cross section area $A$ as $Q^f_0$:

$$Q^f_0 = \alpha A^2 \frac{1}{y^4-x}, \qquad (B1)$$

The flow rate in every of the Sierpinski fractals which fill the tubes added at iteration $n$ is written as follows:

$$Q^f_n = Q^f_0 \left(\frac{1}{y}\right)^{4n} \qquad (B2)$$

and the flow rate in the tubes which are added at iteration $n$ is written as follows:

$$Q_n = Q_0 \left(\frac{1}{y}\right)^{4n}, \text{ where } Q_0 = \alpha A^2 \qquad (B3)$$

To construct the double fractal, at every iteration of the original Sierpinski prefractal, the new tubes are filled with smaller Sierpinski fractals. The flow rate in the prefractal $Q^F_n$ can be computed as a function of the iteration number as in the following recursive formula:

$$Q^F_n = Q^F_{n-1} - x^{n-1} Q_n + x^{n-1} Q^f_n \qquad (B4)$$

The second term corresponds to the decrease of flow rate due to blocking $x^{n-1}$ tubes and the third term depicts the increase of flow rate when the blocked tubes are filled with the small fractals. The recursion in B4 can be rewritten using the notation introduced in Eq. B1-B3:

$$Q^F_n = Q^f_0 - \sum_{k=1}^{n} Q_k x^{k-1} + \sum_{k=1}^{n} Q^f_k x^{k-1} \qquad (B5)$$

Substitution of Eq.B1-B3 into Eq. B4 yields:

$$Q_n^F = Q_0^f - Q_0 \sum_{k=1}^{n}(\frac{1}{y})^{4k} x^{k-1} + Q_0^f \sum_{k=1}^{n}(\frac{1}{y})^{4k} x^{k-1} \tag{B6}$$

Subsequently, the permeability of the double fractal $Q^F$ is computed as a result of an infinite number of iterations:

$$Q^F = Q_0^f - Q_0 \sum_{k=1}^{\infty}(\frac{1}{y})^{4k} x^{k-1} + Q_0^f \sum_{k=1}^{\infty}(\frac{1}{y})^{4k} x^{k-1} \tag{B7}$$

This equation can be written in terms of the dimensionless permeability defined in Eq. 15:

$$K^F = K_f - \sum_{k=1}^{\infty}(\frac{1}{y})^{4k} x^{k-1} + K_f \sum_{k=1}^{\infty}(\frac{1}{y})^{4k} x^{k-1} \tag{B8}$$

Note that

$$\sum_{k=1}^{\infty}(\frac{1}{y})^{4k} x^{k-1} = \frac{1}{y^4}\sum_{k=1}^{\infty}(\frac{1}{y})^{4(k-1)} x^{k-1} = \frac{1}{y^4}\sum_{k=0}^{\infty}(\frac{x}{y^4})^k = K_f \tag{B9}$$

Thus, Eq. B8 is reduced to following:

$$K^F = K_f - K_f + K_f K_f = K_f^2 \tag{B10}$$

In order to compute the permeability of a bundle where the Sierpinski triangular carpets fill the Pascal triangle (we assume that both fractals are composed of equilateral triangles), Eq. B6 and Eq. B7 should be re-written as:

$$Q_n^F = {}^P Q_0^f - {}^P Q_0 z_P \sum_{k=1}^{n}(\frac{1}{y_P})^{4k} x_P^{k-1} + {}^S Q_0^f z_P \sum_{k=1}^{n}(\frac{1}{y_P})^{4k} x_P^{k-1} \tag{B11}$$

$$Q^F = {}^P Q_0^f - {}^P Q_0 \frac{z_P}{y_P^4 - x_P} + {}^S Q_0^f \frac{z_P}{y_P^4 - x_P}, \tag{B12}$$

where the additional subscripts *P* and *S* refer to the Pascal triangular fractal and the Sierpinski carpet respectively. If the Pascal fractals fill the Sierpinski triangular carpet, Eq. B12 takes the form:

$$Q^F = {}^SQ_0^f - {}^SQ_0 \frac{1}{y_S^4 - x_S} + {}^PQ_0^f \frac{1}{y_S^4 - x_S} \tag{B13}$$

If the expressions for flow rate in Eq. B12-B13 are written using the dimensionless permeability, the two first terms cancel out and the resulting permeability in both cases reads:

$$K^F = K^P K^S = \frac{z_P}{y_P^4 - x_P} \frac{1}{y_S^4 - x_S} \tag{B14}$$

C) Permeability-porosity relationships for double prefractal bundle generated with the Pinwheel fractal

The permeability of this prefractal is computed as the partial sum of the series shown in Eq. B7:

$$K(n) = \frac{1}{y^4} \sum_{k=0}^{n-1} \left(\frac{x}{y^4}\right)^k - \sum_{k=1}^{n} \left(\frac{1}{y^4}\right)^k x^{k-1} + \frac{1}{y^4} \sum_{k=1}^{n} \left(\frac{1}{y^4}\right)^k x^{k-1} \sum_{l=0}^{n-1} \left(\frac{x}{y^4}\right)^l \tag{C1}$$

The formula for the porosity $\varphi(n)$ as a function of the iteration number *n* can be obtained using Eq. C1 where $y^4$ is replaced by $y^2$. The size distribution function can also be calculated utilizing the same equation but assuming $y=1$ in all terms of the formula. The partial sums of the geometric series in Eq. C1 can be simplified noting that

$$\sum_{k=1}^{n} \left(\frac{1}{y}\right)^{4k} x^{k-1} = \frac{1}{y^4} \sum_{k=0}^{n-1} \left(\frac{x}{y^4}\right)^k \tag{C2}$$

The simplification of Eq. C1 leads to the following formula:

$$K = \frac{1}{y^4}\left(\frac{x}{y^4}\right)^n + \frac{1}{y^8}\left(\frac{1-(x/y^4)^n}{1-x/y^4}\right)^2 \tag{C3}$$

A formula similar to Eq. C3 describes the porosity of the prefractal bundle as a function of the iteration index *n*:

$$\varphi = \frac{1}{y^2}\left(\frac{x}{y^2}\right)^n + \frac{1}{y^4}\left(\frac{1-(x/y^2)^n}{1-x/y^2}\right)^2 \tag{C4}$$

The relation between porosity and permeability is obtained by eliminating *n* from the two above equations. These equations are quadratic with respect to $(x/y^4)^n$ and $(x/y^2)^n$. The solutions of the quadratic equations read:

$$\left(\frac{x}{y^4}\right)^n = a_1 + a_2\sqrt{a_3 + a_4 K} \tag{C5}$$

$$\left(\frac{x}{y^2}\right)^n = b_1 + b_2\sqrt{b_3 + b_4 \varphi}, \tag{C6}$$

where $a_i$ and $b_i$ are rational functions of the scaling factors *x* and *y*: $a_1 = x^2/2y^4 - 1 - x + y^4/2$; $a_2 = -x/2y^4 + 1/2$; $a_3 = x^2 - 4y^4 - 2xy^4 + y^8$; $a_4 = 4y^8$; $b_1 = x^2/2y^2 - 1 - x + y^2/2$; $b_2 = -x/2y^2 + 1/2$; $b_3 = x^2 - 4y^2 - 2xy^2 + y^4$; $b_4 = 4y^4$.

In the left part of the equations, the scaling factor *x* can be substituted by $x = y^D$ and then the iteration index *n* can be eliminated yielding

$$K = c_1 + c_2\left(c_3 + c_4\left(b_1 + b_2\sqrt{b_3 + b_4\varphi}\right)^{\frac{4-D}{2-D}}\right)^2, \tag{C7}$$

where $b_i$, and $c_i$, $i=1\text{-}4$ are rational functions of the scaling factors *x* and *y*.

D) Heat transfer in prefractal tube bundles

Modeling of flow in porous media often includes a number of physicochemical transport processes such as mass or heat transfer. If the porous medium has irregular or complex hierarchical structure, the fractal approach is successfully used for analysis of the transport processes (Perfect and Sukop 2001, West et al. 1997). For the fractal tube bundles discussed above, the main parameters of heat/mass transfer in the porous medium can be written as a function of the porosity in a similar manner as the relationships between permeability and porosity shown above.

The derivation of the equation for an additional transport process can be exemplified by the relationships between porosity and an average bundle Nusselt number assuming that the heat transfer occurs at a constant temperature $T_w$ of the walls of bundle tubes and for a given inlet fluid temperature $T_0$. The Nusselt number for an individual tube is calculated as $Nu=hd/k_h$ where $h$ is the heat transfer coefficient and $k_h$ is the thermal conductivity of the fluid. For fully developed laminar flow, the Nusselt number is a constant (see various examples in Shah 1978) which we denote by $Nu^*$. The heat flux $j$ in the tube is written as

$j=k_h\ Nu^*(T_w\text{-}T_0)/d$ (D1)

and the total heat flux $J$ at the walls of all tubes in the bundle is calculated as:

$$J = k_h\ Nu_*(T_w - T_0)m_{-1} = \int_d^{d_0} n(\xi)j(\xi)d\xi = k_h\ Nu_*(T_w - T_0)\int_d^{d_0} n(\xi)/\xi d\xi$$

(D2)

If a Nusselt number $Nu_B$ of the fractal bundle is defined as: $Nu_B=J\ d(k_h\ (T_w\text{-}T_0))^{-1}$, the relationships between the bundle Nusselt number and the porosity can be written using the power law pore size distribution as for Eq.8:

$$Nu_B = c_1(1-\varphi)^{-\frac{1+D}{2-D}} - c_2,$$ (D3)

where the coefficients are calculated as:

$$c_1 = \frac{Cd}{(D+1)} Nu^* \left[\frac{C\pi}{4(2-D)A}\right]^{(1+D)/(2-D)} \text{ and } c_2 = \frac{Cd}{(D+1)} Nu^* d_0^{-D-1} \quad \text{(D4)}$$

Since the surface area of the prefractal increases as a function of porosity, the exponent in Eq. D3 is negative, and the bundle Nusselt number $Nu_B$ tends to infinity when the porosity approaches unity. The form of the exponent is defined by the relationship between second positive (Eq. 2) and first negative (Eq. D2) moments of the distribution density.

The possibility to explicitly calculate the moments of the distribution density specified by a power law could also be utilized for the derivation of the Taylor dispersion coefficient in the prefractal tube bundles. For example, the correction factors of the Taylor dispersion coefficient in such prefractals could be obtained following the approach suggested by Carbonell (1979) for the analysis of Taylor diffusion in tube bundles.

E) Regression coefficients in Fig. 5

1) $K_f$=1.40306; $K_*$=56.0927; D=1.83418e-15; $R^2$=0.998324; rms=0.0468576

2) $K_f$=2.63448; $K_*$=105.217; D=1.57839e-15; $R^2$=0.997981; rms=0.0997096

3) p=0.618854; r=0.234571; $R^2$=0.996432; rms=0.0683661

4) p=1.084650; r=0.256929; $R^2$=0.999578; rms=0.0458130

Note that the fitting values of parameter D are less than unity while the fractal dimension in Eq. 21 has to be greater than one. Except the apparent limitations of use of the capillary model for assessment of three-dimensional flow it is necessary to add one more comment about accuracy of the fitting procedure per se. Fitting of the permeability data either with Eq. 21 or with the Archie's law is an ill-conditioned inverse problem in the same sense as for example well known ill-posed problem of fitting of kinetics rates with the Arrhenius equation. Consequently fittings may have non-unique solution or two solutions which are close in some metric but with the values of fitting parameters which are quite distinct from each other. For example, fitting curve for 20 ppi foam is obtained as a result of computation with a Mathematica 7.01.0 script (see Script 1 with results printout below) with three fitting parameters. Fitting procedure can be also executed restricting the least square method to only two parameters while the value of parameter D scans range from one to two with a prescribed step (see Script 2 with results printout below). The latter fitting results in D=1.0201(>1) and in a high value of $R^2$= 0.996999 which is only 0.14% lower than $R^2$ of the former fitting. In the latter case, RMS increases by 0.016 (x10 $^7$ m$^2$ ) that is less than 2% of the measured permeability values. Thus if relative error of the permeability measurements is greater than 2%, it may be impossible to reliably choose between the former and the latter fittings.

```
"Script 1: 20 ppi permeability dataset, LSM with three parameters"
data={{0.9005,0.9},{0.906,.854},{0.9245,1.1},{0.949,1.185},{0.9546,1.3},{0.978,1.42}};

(* LSM with three parameters*)
model2=b2-(bbb2*bbb2)*(1-arg)^((bb2^2-4)/(bb2^2-2));

f2=FindFit[data,model2,{b2,bb2,bbb2},arg,MaxIterations→1000,Gradient→"FiniteDifference"];

nlm1=NonlinearModelFit[data,model2,{b2,bb2,bbb2},arg,MaxIterations→1000,Gradient→"FiniteDifference"];

Print[Row[{"Kf=",b2/.{f2[[1]]}}]];
Print[Row[{"K*=",(bbb2/.{f2[[3]]})^2}]];
Print[Row[{"D=",(bb2/.{f2[[2]]})^2}]];
Print[Row[{"R2=",nlm1["RSquared"]}]];
Print[Row[{"RMS=",RootMeanSquare[nlm1["FitResiduals"]]}]];
```

```
Plot[xb2-(xbbb2*xbbb2)*(1-x)^((xbb2^2-4)/(xbb2^2-
2)),{x,0.9*data[[1,1]],1},Epilog→{PointSize[Medium],Po
int[data]}]
```

 Script 1: 20 ppi permeability dataset, LSM with three
parameters
 Kf= 1.40306
 K*= 56.0927
 D= 1.83418×10⁻¹⁵
 R2= 0.998324
 RMS= 0.0468576

"Script 2: 20 ppi permeability dataset, LSM with two
parameters and D scanning from 1 to 2"
```
data={{0.9005,0.9},{0.906,.854},{0.9245,1.1},{0.949,1.1
85},{0.9546,1.3},{0.978,1.42}};

(*LSM with two parameters and D scanning from 1 to 2"*)
rms0=1000.;xb20=0;xbb20=0;xbbb20=0;r20=0;

For[i=1,i<41,i++,

 bb2=1+0.01*i;

 model2=b2-(bbb2*bbb2)*(1-arg)^((bb2^2-4)/(bb2^2-2));

f2=FindFit[data,model2,{b2,bbb2},arg,MaxIterations→100
0,Gradient→"FiniteDifference"];

nlm1=NonlinearModelFit[data,model2,{b2,bbb2},arg,MaxIte
rations→1000,Gradient→"FiniteDifference"];

 r2=nlm1["RSquared"];
 rms=RootMeanSquare[nlm1["FitResiduals"]];
 xb2=b2/.{f2[[1]]};
 xbbb2=bbb2/.{f2[[2]]};

If[rms<rms0,xbbb20=xbbb2;xbb20=bb2;xb20=xb2;r20=r2;rms0
=rms];

 ](*end For[]*)

Print[Row[{"Kf=",xb20}]];
Print[Row[{"K*=",xbbb20^2}]];
Print[Row[{"D=",xbb20^2}]];
Print[Row[{"R2=",r20}]];
Print[Row[{"RMS=",rms0}]];
```

```
Plot[xb20-(xbbb20*xbbb20)*(1-x)^((xbb20^2-4)/(xbb20^2-
2)),{x,0.9*data[[1,1]],1},Epilog→{PointSize[Medium],Po
int[data]}]
```
```
 Script 2: 20 ppi permeability dataset, LSM with two
parameters and D scanning from 1 to 2
 Kf= 1.33731
 K*= 563.178
 D= 1.0201
 R2= 0.996999
 RMS= 0.0627012
```

F) Visualization of mesh with different porosity

A possible explanation of the qualitatively similar behavior of the permeability of the metal foams and the prefractal bundles can found in the following example: let us assume that a sample of a solid mesh is inserted into a tube and is a subject to a flow. If the sample is relatively small and the mesh pores are relatively large, some space will be available for flow through the sample along straight flow lines. This space can be visualized by a projection of the mesh on the plane perpendicular to the tube axis. As an example of such mesh, Fig. 6 shows a torus that is defined in the library of graphic objects in software package Mathematica (http://www.wolfram.com/). The images in Fig. Appendix F a) and b) are plotted setting different line thickness in order to imitate different thickness of the mesh wire and consequently porous media with different porosity. Despite the fact that the mesh is composed of regularly spaced lines, the overlapping of the mesh cells for the chosen projection leads to a relatively wide distribution of space opened for straight (parallel to the axis of symmetry) flow lines. In order to assess the pore size distribution function of the open space and two-dimensional porosity, the images shown in Fig. 6 are binarized and then analyzed applying the box counting algorithm which is implemented in the ImagJ software package (ImagJ http://rsbweb.nih.gov/ij/ ). The results show that the porosity of the images differs by 70% while the distribution functions of the both images are well ($R^2>0.9$) approximated by the power law distribution with the same exponent coefficient of 1.6. The example suggests that the space available for flow along straight or slightly curved flow lines in high-porosity metal foams may have the power law distribution function and thus the permeability-porosity relationships of such foams could be represented by the formulas derived herein for the fractal bundles.

Figures

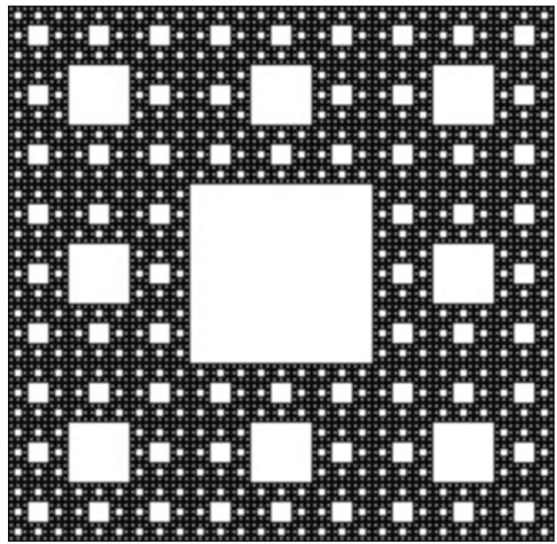

Fig.1 Square Sierpinski carpet (as shown in wikipedia.org/wiki/List_of_fractals_by_Hausdorff_dimension)

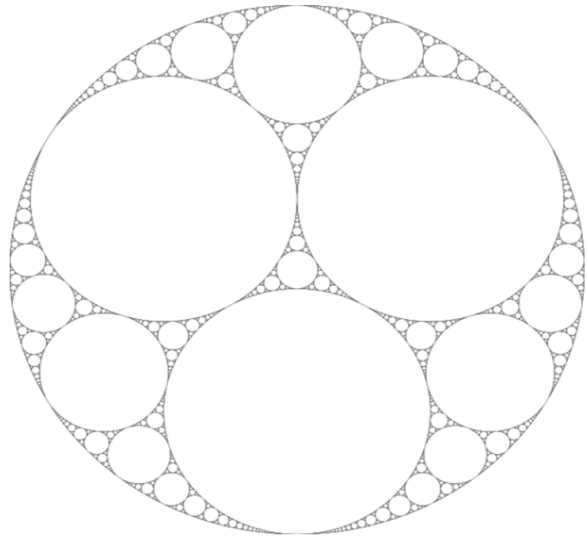

Fig.2. Apollonian gasket (as shown in wikipedia.org/wiki/List_of_fractals_by_Hausdorff_dimension)

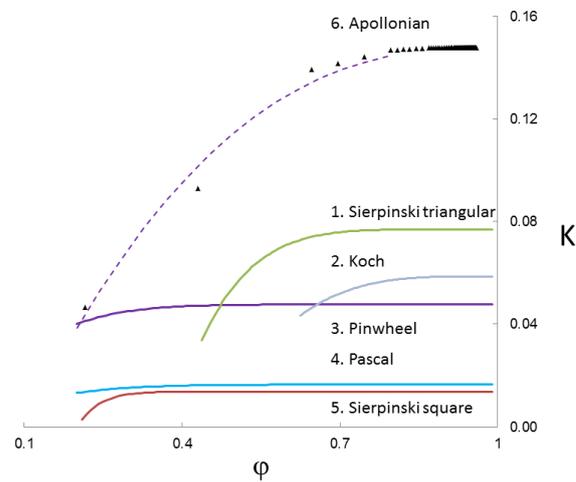

Fig. 3 Dimensionless permeability of geometric prefractals: 1) Sierpinski triangular carpet, D=1.59; 2) Koch snowflake, D=1.26; 3) Pinwheel fractal, D=1.73; 4) Pascal triangular fractal, D=1.68; 5) Sierpinski square carpet, D=1.89;

Apollonian gasket D=1.3057: data (closed triangles) and fitting curve (dashed line 6).

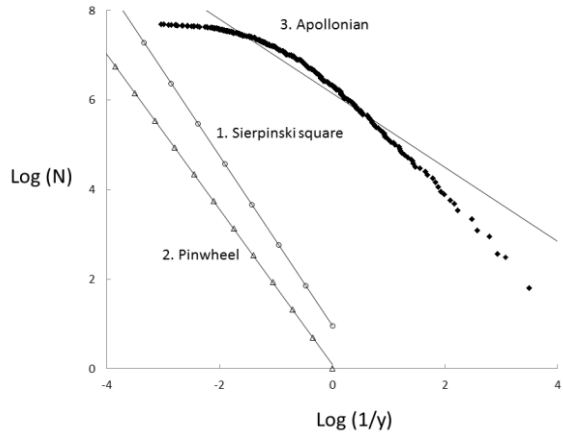

Fig. 4 Size distribution functions. 1) Sierpinski square carpet, D=1.89. Symbols: Eq.16, solid line: the least square fit; 2) Pinwheel fractal, D=1.73. Symbols: Eq.16, solid line: the least square fit; 3) Apollonian gasket D=1.3057. Symbols: Jacquenot, 2007, solid line: the least square fit with a power law equation (exponent coefficient 0.83)

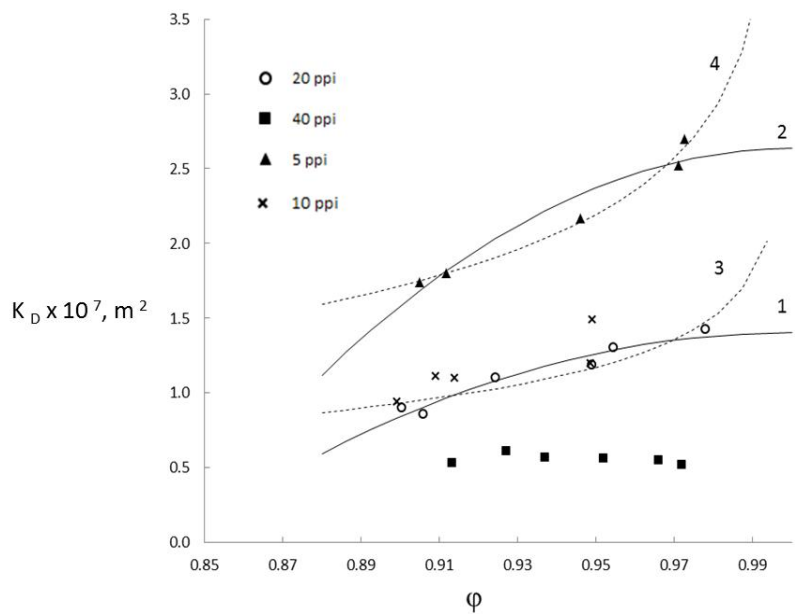

Fig. 5 Permeability of metal foams adopted from table 2 in (Bhattacharya et al. 2002); legend indicates number of pores per inch (ppi) in tested foam; 1,2 – fitting curves (20 and 5 ppi) with Eq. 21; 3,4- empirical correlation (20 and 5 ppi) of Rodriguez et al. 2004.

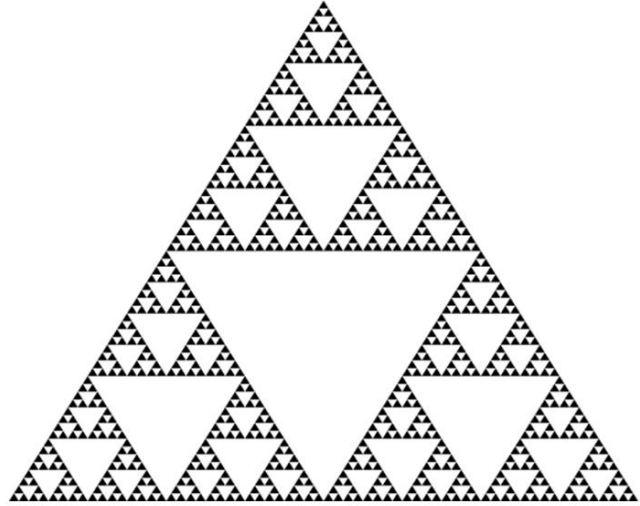

Fig. A1 Sierpinski triangular carpet (as shown in wikipedia.org/wiki/List_of_fractals_by_Hausdorff_dimension)

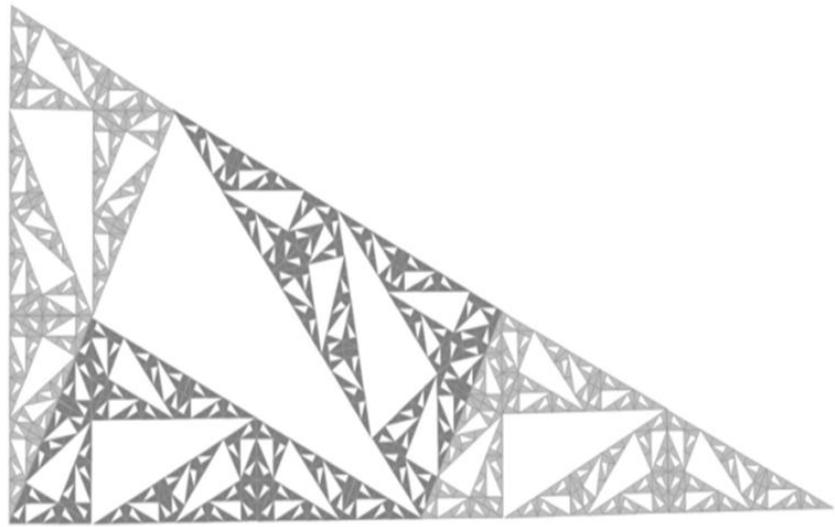

Fig. A2 Pinwheel fractal (as shown in wikipedia.org/wiki/List_of_fractals_by_Hausdorff_dimension)

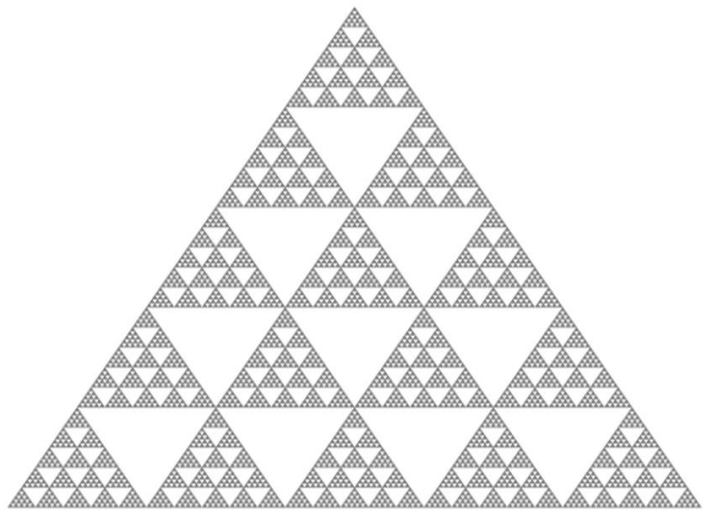

Fig. A3 Pascal triangle modulo 5 fractal (as shown in wikipedia.org/wiki/List_of_fractals_by_Hausdorff_dimension)

.

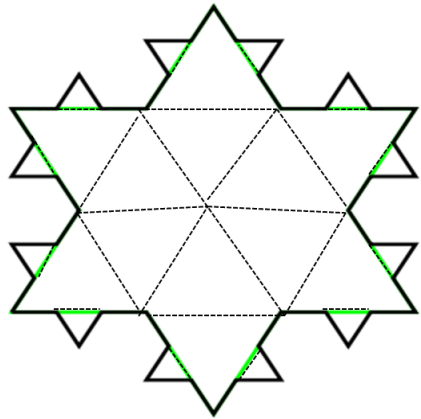

Fig. A4. Koch tube bundle after the second iteration

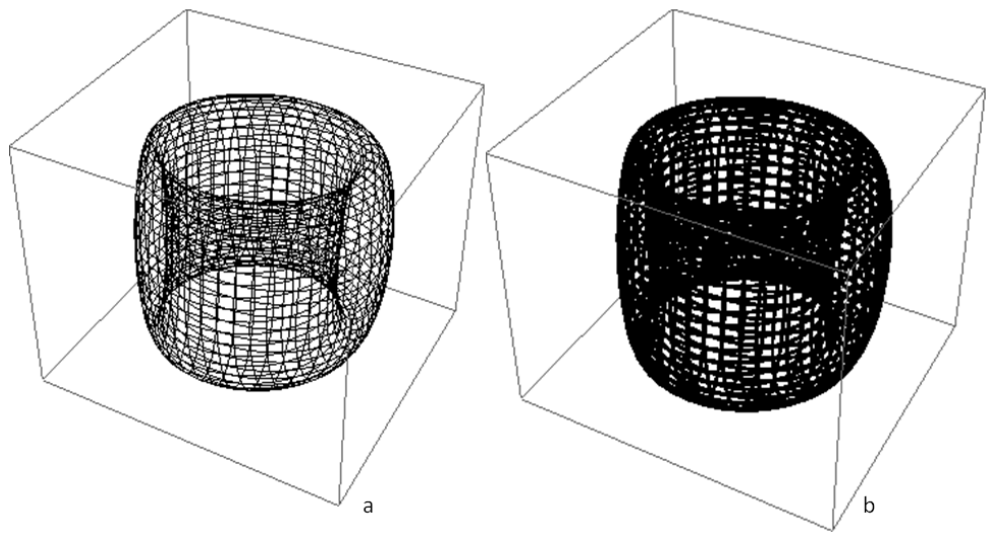

Fig. F1. Mesh on the graphics object Torus from the library of the Mathematica (http://www.wolfram.com/) software package: a) line thickness 0.001; b) line thickness 0.006.